\documentclass{amsart}
\usepackage{amsaddr}
\usepackage{amsthm}
\newtheorem{theorem}{Theorem}
\usepackage{etex}
\usepackage[utf8]{inputenc}
\usepackage[binary-units]{siunitx}
\usepackage{xkvltxp}
\usepackage[usenames]{xcolor}
\usepackage[status=draft]{fixme}
\fxusetheme{color}
\renewcommand{\fxnote}[1]{\relax}

\newtheorem{observation}[theorem]{Observation}
\theoremstyle{definition}
\newtheorem{definition}{Definition}

\usepackage{amsmath}
\usepackage{amssymb}
\usepackage{pgfplots}
\usepackage{pgfplotstable}

\pgfplotstableread{
Size	1	2	3	4	5
1	0.825697536	13.72517713	3.975114427	0.7824942	6.877500157
2	3.036627512	8.243970113	3.383952987	0.609867172	3.368506058
3	7.527479198	20.27881703	7.553675	1.235270635	8.086040343
4	0.871685334	14.49920774	5.203223401	0.830852221	10.43929909
5	2.439854181	10.00441095	2.530418824	0.892103517	3.228100498
6	1.098490359	16.58409758	6.527760862	0.84765545	11.85874009
7	0.713835082	11.4319845	3.425400397	0.814868746	10.45690592
}\mytable

\pgfplotstableread{
Size	1	2
1	42.23218	47.679782
2	21.344757 32.738376
4	11.288476	18.68422
8	5.986905	18.799996
12	4.034853	19.485068
16	3.091908	19.04273
20	2.586699	19.466782
}\mytables
\usepackage{filecontents}

\newcommand{\Z}{\mathbf{Z}}
\newcommand{\R}{\mathbf{R}}
\renewcommand{\vec}[1]{{\mathbf{#1}}}
\newcommand{\ve}[1]{{\vec{#1}}}
\newcommand{\mat}[1]{{\mathbf{#1}}}
\newcommand{\ma}[1]{{\mat{#1}}}
\makeatletter
\def\imod#1{\allowbreak\mkern10mu({\operator@font mod}\,\,#1)}
\makeatother

\usepackage{booktabs}
\usepackage[singlelinecheck=off]{caption}
\usepackage{subfigure}
\usepackage{tikz}
\usetikzlibrary{shapes,decorations,calc,external}
\usepackage[ruled,norelsize]{algorithm2e}

\SetAlFnt{\small}
\SetAlCapFnt{\small}
\SetAlCapNameFnt{\small}
\SetAlCapHSkip{0pt}
\IncMargin{-\parindent}

\begin{document}

\title{Model-Driven\\ Automatic Tiling with\\ Cache Associativity Lattices}
\author{David Adjiashvili}
\address{IFOR, D-Math, ETH Zürich, Rämistrasse 101, 8092 Zürich, Switzerland}

\author{Utz-Uwe Haus}
\address{Cray EMEA Research Lab, Hochbergerstrasse 60C, 4057 Basel, Switzerland\\ (current address; previously IFOR, D-Math, ETH Zürich)}

\author{Adrian Tate}
\address{Cray EMEA Research Lab, Hochbergerstrasse 60C, 4057 Basel, Switzerland}

\begin{abstract}
Traditional compiler optimization theory distinguishes three separate classes of cache miss -- Cold, Conflict and Capacity. Tiling for cache is typically guided by capacity miss counts. Models of cache function have not been effectively used to guide cache tiling optimizations due to model error and expense. Instead, heuristic or empirical approaches are used to select tilings. We argue that conflict misses, traditionally neglected or seen as a small constant effect, are the only fundamentally important cache miss category, that they form a solid basis by which caches can become modellable, and that models leaning on cache associatvity analysis can be used to generate cache performant tilings. We develop a mathematical framework that expresses potential and actual cache misses in associative caches using \emph{Associativity Lattices}. We show these lattices to possess two theoretical advantages over rectangular tiles -- volume maximization and miss regularity. We also show that to generate such lattice tiles requires, unlike rectangular tiling, no explicit, expensive lattice point counting. We also describe an implementation of our lattice tiling approach, show that it can be used to give speedups of over 10x versus unoptimized code, and despite currently only tiling for one level of cache, can already be competitive with the aggressive compiler optimizations used in general purposes compares such as GCC and Intel's ICC. We also show that the tiling approach can lead to reasonable automatic parallelism when compared to existing auto-threading compilers. 
\end{abstract}

%
%







\maketitle

\section{Introduction}

Tiling has been shown to be a robust and effective transformation for
exploiting locality~\cite{lam1991cache} and for
parallelism~\cite{anderson1993global}. Though general purpose
compilers perform some tiling for cache performance, they also tend to
rely on heuristic approaches which when lacking information will act
conservatively. In the Polyhedral Model~\cite{bastoul-habil:12}, a lot
of progress has been made regarding how to generate tiled
codes~\cite{Ramanujam1992108} and how the tiles should be chosen to
exploit locality and to extract
parallelism~\cite{bondhugula2008practical}. Even in light of that
research, the so called tile-size selection problem remains open,
leading some researchers to suggest that only empirical auto-tuning
can provide the answers~\cite{chen2008chill}. Our research takes a
very different approach: at the philosophical level we believe that
advances in understanding and modeling of memory hierarchies remain
unsolved, and that with greater understanding models of memory and
cache can be used to guide tiling algorithms. In particular, we
believe that associativity in cache architectures is the fundamentally
important though much neglected feature of cache memories, and when
better understood can lead to the generation of more accurate cache
models. We will first make this argument informally while describing
the features of associative cache in Section~\ref{sec:cache}. In
Section~\ref{sec:cache-miss-machinery} we will then begin to describe
the modeling framework to provide first a working model of potential
cache misses in Section~\ref{sec:potential-misses}, and later a
working model of actual cache misses in
Section~\ref{sec:actual-misses}. Then in
Section~\ref{sec:tiling-model} we will describe how this model extends
to tiled codes. We will describe a tiling framework and a simulation
environment that, taking a specification as input, can build the
appropriate cache model for the operation, choose the tiling that
minimizes cache misses for a single level of the hierarchy, and then
generate the appropriate tiled codes. We will describe performance
results of this framework in Section~\ref{sec:experimental}. Our work
is highly experimental and could be built-upon on many different
ways. We discuss those ways and related research in
Section~\ref{sec:related-work}.

\subsection{Associative Cache Function}\label{sec:cache}

\subsubsection{Cache Specification}

\fxnote{maybe a figure here --da}
In order to fully describe the need for tiling according to
associativity, we first describe the mechanism of $K$-way
set-associative caches. Data is moved in units called
\emph{lines}. All addresses map to several \emph{memory-level-lines}
such as \emph{memory-line} (meaning a line in DRAM), \emph{L1
  cacheline} and \emph{L2 cacheline}. When data residing at a memory
address needs to be accessed, the caches are first checked for the
presence of the corresponding line, and it is loaded from the highest
(i.e., closest) cache level in which the line is present, at minimum
access cost. For a specific cache level, each memory address is mapped
deterministically to a certain cache \emph{set} containing $K$
possible cache slots or \emph{ways}. The cache-level functionality can
thus be specified by a \emph{cache specification} $C=(c,l,K,\rho)$,
where $c$ is the total cache capacity (total bytes that can be stored
in the cache), $l$ the cache line size (number of bytes fetched in one
load), $K$ the associativity (number of cachelines that can reside in
one cache set), and $\rho$ an index $\rho=1,\ldots,P$ (the cache's
position in a $P$-level memory hierarchy). Such a specified cache has
$N=\tfrac{c}{lK}$ cache sets, and hence every $(\tfrac{c}{lK})^{th}$
cacheline or $(\tfrac{c}{K})^{th}$ data element maps to the same
set. This simple striding defines the mathematical structure on which
our cache models are based.

\subsubsection{Cache Miss}

When data at a given address must be used, and the associated line is
not found in the cache, then a \emph{cache miss} occurs and an
expensive load from a lower memory level ensues. The literature has
traditionally differentiated between three categories of cache misses:
\emph{capacity misses} where data needs to be loaded because more
cachelines are accessed than can fit into total cache, \emph{cold
  misses} where data has never been accessed and thus must be loaded,
and \emph{conflict misses} where a line must be loaded because,
although previously in the cache, the line was evicted when too many
cachelines were loaded into the same cache set. Typically,
associativity is considered a small constant
effect~\cite{hennessy2012computer} and is ignored by most cache models~\cite{yotov2005search,cascaval-padua:03}. We believe that the effects of
associativity have been misunderstood and neglected, and further that
cache misses in associative caches are better categorized using a
single classification: that of conflict misses due to
associativity. We justify this informally here and develop it formally
in Section~\ref{sec:actual-misses}.

\subsubsection{Cache Capacity}

\begin{figure}
  \centering
  \includegraphics[width=0.4\textwidth]{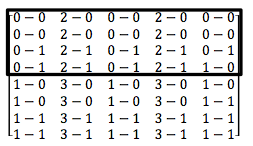}
  \caption{An $8\times5$ 2-d array stored in column-major order with
    cachelines of length 2, and where each data element is marked
    Set-Line (e.g. 1-0 maps to set 1 line 0). If we attempt to address
    only the upper $2\times5$ sub-array (bordered) and loaded this
    into a 2-way associative cache with 4 sets, then it is not
    possible to address the sub-array without cache misses, since the
    sub-array contains three cachelines that map to sets 0 and 2.}
  \label{fig:capacity}
\end{figure}

We first note that cache capacity, though perhaps occasionally a
useful approximation for programmers, is neither expressed nor
comprehended in cache logic, and can lead to misleading and inaccurate
estimations of the data volume accessible to a cache, when accessing
tiles or padded array segments. Cache protocols assume the perspective
of a single cache set. Since all data map to a given cacheline and all
cachelines map to a given set, then only a single set is checked for
the presence of the cacheline in question. Correspondingly, only the
contents of one set are candidates for eviction. Only in the case
where all cache sets are used uniformly does total cache capacity
remain a useful quality. Any variation in usage between sets (which is
typical) decreases the accuracy of cache capacity as a metric. The
example in Figure~\ref{fig:capacity} illustrates how a 2-d array
stored in a 2-way associative cache with 4 sets cannot use full cache
capacity. The extreme of this effect is called \emph{cache thrashing},
where consecutively accessed elements map to identical sets, and can
be seen as the lower bound of cache capacity usefulness. Since the
single measure of cache capacity is variable it does not serve as a
suitable model parameter. The cache capacity per set does remain
valid, though it should be obvious that per-set cache capacity is treated by 
the \emph{conflict miss} category of misses.

Cold misses also do not require any special treatment and can be
viewed as a special case of conflict miss. In
Section~\ref{sec:actual-misses} we will show that a given set
of~\emph{potential cache misses}, meaning a group of cachelines that
all map to the same set, can be further categorized as \emph{actual
  cache misses} when the reuse distance between successive reuses
exceed the cache associativity. In a typical set, the situations that
produce this will be when either the cacheline has never been used
before (cold misses) or when more than $K$ different cachelines have
been used before the reuse (conflict misses). We therefore choose to
model cache through the single mechanism of associativity misses and
will refer to these as cache-misses. This informal reasoning is made
concrete in Section~\ref{sec:actual-misses}.

\subsubsection{Cache Reuse Policy}

The exact mechanism used by a set to decide if a cacheline should be
evicted, and which line to evict, is called the \emph{eviction
  policy}. We will consider two evection policies in this paper, which
are the most commonly implemented reuse policies in modern hardware --
\emph{Least Recently Used (LRU)} and \emph{Pseudo Least-Recently Used
  (PLRU)}. Our framework implements model variations for both
policies. Our implementation allows us to see a comparison of these
reuse policies and to see which policy appears to match experimental
results more closely (and is therefore more likely implemented in the
hardware). A detailed description of the policy and the effects that
policy choice has on model quality is interesting but deferred to a
future paper.

\subsection{Related Work}\label{sec:related-work}
Ghosh et al.~\cite{ghosh1997cache} describe a mathematical framework
for the evaluation of cache misses. Solutions to their Cache Miss
Equations (CMEs) correspond directly to the cache misses in a
code. However, as pointed out in their work, the CMEs are ultimately
intractable. For certain situations, the CMEs can allow users to
understand a lot about cache behavior because basic number theory can
be used to describe situations when no cache misses arise. They cannot
however be used to model accurately general situations when misses do
arise. The base mathematical property of the CMEs is the same as in
our work. However their work did not realize the inherent lattice
structure that is formed by the solution of the CMEs.  \fxnote{Must
  position against Abella2002. Their approach is inferior but the
  package more complete. Currently we present no case for who the cost
  of analysis can be incorporated into a compiler or DSL}

CMEs were used in~\cite{abella2002near-optimal} to drive tiling
transformations and a reduction in the ratio of capacity to conflict misses for several benchmarks was shown. However absolute performance of the tilings was not discussed, and the research was continued in a meaningful way. While CMEs may be useful in isolated cases, their ultimately untractable solution space means that their applicability for general transformations is unlikely.

We have expressed cache misses in a concise mathematical formulation for which we believe future work can yield efficient optimal or approximate code generating schemes. To a limited extent, this is already evident in our work. The lattice tiles can be generated quite simply without counting of lattice points, and a relatively simple decision algorithm can be incorporated into the model-based decision-making. However, the model as currently expressed is non-polynomial in execution time.
No research to date has constructed tiles based on the associativity characteristics of a memory. \cite{goumas-athanasaki-koziris:03} contains analysis of various rectangular tiles and observes that lattice tilings would be theoretically superior, but to our knowledge this lead was not followed by the authors or any other researchers. 

\section{Basic Cache Miss Model}

Throughout the discusion we will assume a cache specification
$C=(c,l,K,\rho)$ is given, and we denote by $N=\tfrac{c}{lk}$ the
number of cache sets in it.

\subsection{Cache Miss Machinery}\label{sec:cache-miss-machinery}
\subsubsection{Index Maps}

Let $A$ be a $(m_1,\dots,m_d)$-table, w.l.o.g with index set
$Q(A)=[0,m_1-1]\times\dots\times[0,m_d-1]\cap\Z^d$. In RAM the
elements of $A$ will be spread out into a (typically consecutive)
1-dimensional array which we will denote by $a(A)$ (or simply $a$) of
size $m_1\cdot\ldots\cdot m_d$. Its elements are $a_i$ for
$i\in\{0,\dots,m_1\cdot\ldots\cdot m_d-1\}$.

\begin{definition}[index map]
  Given a $(m_1,\dots,m_d)$-table $A$ with array $a(A)$, a bijective
  function
  $$\phi:Q(A)=[0,m_1-1]\times\dots\times[0,m_d-1]\cap\Z^d\to a(A)=[0,m_1\cdot\ldots\cdot
  m_d-1]\cap\Z$$ is called an \emph{index map} for the pair
  $(A,a)$. Its inverse is $\phi^{-1}$.
\end{definition}

Typical index maps are affine functions like
$$\phi_c(i_1,\dots,i_d)=i_1+m_1(i_2+m_2(i_3+m_3(\dots+m_{d-1}i_d)\cdots))=\sum_{k=1}^d(\prod_{l=1}^{k-1}m_l)i_k$$
(column major order) or
$$\phi_r(i_1,\dots,i_d)=i_d+m_d(i_{d-1}+m_{d-1}(i_{d-2}+m_{d-2}(\dots+m_{2}i_1)\cdots))=\sum_{k=1}^d(\prod_{l={k+1}}^{d}m_l)i_k$$
(row major order). Their inverses can be defined using $\mathrm{mod}$ and
$\mathrm{div}$.

Since $a(A)$ is naturally ordered, for each index map $\phi$ there
exists a unique point $q_A\in Q(A)$ such that $\phi(q)$ has
minimal index in $a(A)$ such that $\phi(q_A)=0\imod N$. This point
will be called the \emph{base point} of $Q(A)$.

We will mostly consider affine index maps, i.e.
$\phi(x+y)=\phi(x)+\phi(y)$ and $\phi(\lambda x)=\lambda\phi(x)$. In
this case $\phi(q_A)$ is exactly the affine offset. We can furthermore
assume that $\phi$ is monotone wrt. the component-wise ordering of $Q$
(otherwise we need to consider, e.g.,
$\phi'(x_1,\dots,x_d)=\phi(x_1,\dots,m_i-x_i-1,\dots,x_d)$. If
$q_A=0\in Q(A)$ we will sometimes say that $\phi$ is linear (which it
is, as a map to the module $\Z/N\Z$). Non-linear index maps, like
sparse matrix storage using auxiliary mapping arrays, are also
interesting, but beyond the scope of this paper.

\subsubsection{Iteration Domains}

Let two tables $A$ and $B$ be given. When we consider an arbitrary
pair of elements $x\in A$ and $y\in B$ we actually index a single
entry in $A\times B$. For typical computations, like matrix
multiplication we will successively access an entire hyperplane of
$A\times B$ (e.g., to compute $(AB)_{ij}$ we will use indices
$\{(i_1,i_2,i_3,i_4)\in Q(A)\times
Q(B)\::\:i_1=i,i_2=i_3,i_4=j\}$).
Appealing to this use case we will call any affine subspace of such a
product of table index sets an \emph{iteration domain}.

\begin{definition}[iteration domain, operand]
  Given $k$ tables $A_1,\dots,A_k$ with index sets $Q(A_i)\in\Z^{d_i}$
  we call $Q(A_1,\dots,A_k)=Q(A_1)\times\dots\times Q(A_k)$ the
  \emph{joint index set}.\\
  For any affine subspace $H\subseteq\R^{\sum_{i=1}^kd_i}$ the set
  $Q(A_1,\dots,A_k)\cap H$ is
  called a \emph{(joint) iteration domain}.\\
  The tables $A_1,\dots,A_k$ will be called \emph{operands}.  The
  projection function onto operand $i$ will be designated by
  $\pi_i:Q(A_1,\dots,A_k)\to Q(A_i)$.
\end{definition}

Note that $H=\R^{\sum_{i=1}^kd_i}$ is a valid
iteration\fxnote{absolutely must tie PM iteration domain with ours}
domain. Usually the subspace $H$ will be defined so that the iteration
domain remains nonempty. If $H$ is a linear subspace the iteration
domain will be a set of integer points of some sublattice of $\Z^d$
for $d=\sum_{i=1}^kd_i$.
All iteration domains we consider will have a
nonempty affine subspace $H$ in its definition. For typical examples
see Table~\ref{tab:common-subspaces}.

Note that this definition is powerful enough to handle temporal
constraints on iteration: we can add an artificial 1-dimensional
operand whose indices designate the time points, and then add suitable
constraints to the set $H$ to indicate that some combination of
indices of the other operands occurs at multiple time points during
iteration.

\begin{table}\small
  \centering
  \begin{tabular}{lll}
    \toprule
    Operation & algebraic form & constraints\\
    \midrule
    Scalar product        & $A_{0}=\sum_kB_kC_k$ & $\{i_1=0,i_2=i_3\}$\\
    Convolution           & $A_{0}=\sum_kB_kC_{{m^C_1}-k-1}$ & $\{i_1=0,i_2=m^C_1-i_3\}$\\
    Matrix multiplication & $A_{i,j}=\sum_kB_{i,k}C_{k,j}$ & $\{i_1=i,i_2=j,i_3=i,i_4=i_5,i_6=j\}$\\
    Kronecker product     & $A_{m^C_1(i-1)+k,m^C_2(j-1)+l}=B_{i,j}C_{k,l}$ &
    $\{i_1=m^C_1(i_3-1)+i_5,i_2=m^C_2(i_4-1)+i_6\}$\\
    \bottomrule
  \end{tabular}
  \caption{Examples of commonly occuring subspaces. Notation: Table $T$ has 
    index set $Q(T)=\prod_{j=1}^{d^T}[0,m^T_j-1]$.
  }
  \label{tab:common-subspaces}
\end{table}

Traditional reuse analysis is based on reuse vectors and reuse
distances. This concept is not sufficient for high-dimensional
iteration domains where a single vector cannot describe the full reuse
potential. We instead define the \emph{reuse domain} for a given data element
of any operand.

\begin{definition}[reuse domain]
  Let the $k$ operands $A_1,\dots,A_k$ give rise to the iteration
  domain $Q(A_1,\dots,A_k)\cap H$. For a given index of any operand
  $q \in Q({A_i})$, the reuse domain
  $\mathcal{R}_i(q)$ is given by
  \begin{equation*}\label{eq:reuse-domain}
    \begin{aligned}
      \mathcal{R}_i(q) &= Q(A_1) \times Q(A_2) \times \dots \times
      Q(A_{i-1}) \times \{q_i\} \times Q(A_{i+1}) \dots \times Q(A_k) \cap
      H\\
      & = \{x\in Q(A_1,\dots,A_k)\cap H\::\: \pi_i(x)=q_i\}.
    \end{aligned}
  \end{equation*}
  
\end{definition}

Of course a reuse domain can be considered as a particular iteration domain. In fact, is is the best way to iterate over the subset of indices in the joint iteration domain projecting onto $q_i\in Q(A_i)$ to ensure perfect reuse.


\subsection{Ordering and Distance}

Let $D$ be an iteration domain in dimension $d$. 
\begin{definition}[iteration ordering]
Let $\prec \subseteq \Z^d\times\Z^d$ be a total order on $\Z^d$. We will call the restriction $\prec_D$ of $\prec$ to a set $D\subseteq\Z^d$ an \emph{iteration ordering} on $D$.
\end{definition}

Of course, lexicographic ordering of the indices of the tables yields
a iteration ordering, but many other iteration orderings are
conceivable and potentially useful for our application. Furthermore,
an index map $\phi_A$ induces an iteration ordering by virtue of the
natural ordering of $a(A)$, but we will often consider the case where
the index map order and the iteration order are different.

\begin{definition}[subsequent reuse]
Let $\mathcal{R}_i(q)$ be a reuse domain of an iteration domain $D$. Since $\mathcal{R}_i(q) \subseteq D$ then if $\prec$ is a total order on $D$ that also defines a total order on $\mathcal{R}_i(q_i)$. Hence for any non-boundary\fxnote{define? --da} $\ve x \in D$,  $\exists \ve y \in D$ with $ x \prec y $ and $\nexists \ve z \in D$ such that $ x \prec z \wedge z \prec y $. We call $y$ the \emph{subsequent reuse} of $q_i$.
\end{definition}

\begin{definition}[distance]
Given some set $X\subset Z^d$ the set of elements between two points $\vec x\in
X$ and $\vec y\in X$  (including the smaller and excluding the larger of the two) is designated by
$$[\ve x,\ve y) = [\ve x,\ve y)^X_\prec = \{\ve z\in X| \ve x\preceq \ve z \wedge \ve
z\prec \ve y\}.$$
We can thus define a metric on $X$, the distance between $\ve x$ and
$\ve y$ in the order restricted to $X$, as
$\Delta_\prec^X(\ve x,\ve y):=|[\ve x,\ve y)_{\prec}^{X}|+|[\ve y,\ve x)_{\prec}^{X}|$ (at
least one of the the summands will always be $0$ since either
$\ve x\prec_X\ve y$ or vice versa). We call $\Delta_X (\ve x,\ve y)$ the \emph{distance in $X$} between points $\ve x$ and $\ve y$. If $X$ and $\prec$ are clear from context we write $\Delta$ instead of $\Delta^X_\prec$.
\end{definition}

\subsection{Potential Conflicts}\label{sec:potential-misses}

We will first categorize the necessary (but not sufficient) conditions for a cache miss, which we will call a \emph{potential conflict}. Potential conflicts are the set of points in an iteration domain that map to the same set. This notion is independent of both ordering and reuse and says nothing about the actual cache misses that will be incurred -- for this we need to consider orderings (see Section \ref{sec:actual-misses}). First we will define potential conflicts occurring in a single operand, describe the structure of the miss spectrum using a mathematical lattice and then extend both the notion and the structure to define potential conflicts in iteration domains.

\begin{definition}[Operand potential conflicts]
  Let $C=(c,l,K,\rho)$ be a cache specification with $N=\tfrac{c}{lk}$
  cache sets. Given a $(m_1,\dots,m_d)$-table $A$ with array $a(A)$
  and an index map $\phi$ we say that $a_i$ is \emph{in potential
    conflict} with $a_j$ if $i=j~\imod N$. The notion readily extends
  to all elements of $A$ through use of the index map $\phi$.
\end{definition}

What structure emerges from operand potential conflicts? We will
restrict attention in the following to \emph{affine} and bijective
index maps, i.e. $\phi(i_1,\dots,i_d)=\sum_{r=1}^{d-1}w_ri_r + i_d$.
Consider the points in
$Q=[0,m_1-1]\times\dots\times[0,m_d-1]\cap\Z^d$, and their image under
$\phi$. Then the points of the lattice $N\Z=\{Nz\::\:z\in \Z\}$ in the
interval $I=\{0,\dots,m_1\cdot\ldots\cdot m_d-1\}$ induce a lattice
$L=L(C,\phi)\subseteq \Z^d$ such that $\phi(L)=N\Z$ since $\phi$ is an
affine bijection. (We assume wlog. that $\phi$ maps $0\in Q$ to
$0\in N\Z$, i.e. is actually linear; otherwise consider the affine
translate of $L$ by $q_A$.)

What are the generators of $L(C,\phi)$? They can easily be calculated from
the definitions: For linear $\phi$ the lattice $L(C,\phi)$ is generated by
  $$G=\{x-y\::\: x,y\in Q, \phi(x)=\phi(y)\imod N\}.$$ If $G$ is
  empty, there are no potential conflicts for the operand under $\phi$.  

\begin{observation}
  If $\phi(i_1,\dots,i_d)=\sum_{r=1}^{d-1}w_ri_r + i_d$ is an affine
  bijective index map for table $A$ under cache specification
  $C=(c,l,K,\rho)$, then $A_{i_1,\dots,i_d}$ is in potential conflict
  with $A_{j_1,\dots,j_n}$ if and only if they are equivalent modulo
  $L(C,\phi)$, i.e. if $A_{i_1,\dots,i_d} = A_{j_1,\dots,j_n} + l$ for
  some $l\in L(C,\phi)$.
\end{observation}

We will now extend this notion to describe potential conflicts for
iteration domains. The notion extends easily to multiple operands as
follows if we consider conflicts in memory. Let $A_1,\dots,A_k$ be
tables with index maps $\phi_{A_i}$ and joint iteration domain
$Q=Q(A_1,\dots,A_k)$. We say that two entries $p=(p_1,\dots,p_k)\in Q$
and $q=(q_1,\dots,q_k)\in Q$ are \emph{in potential conflict} if the
following condition holds:
  $$\exists i\exists j:\phi_{A_i}(p_i)=\phi_{A_j}(q_j)\imod N$$

In order to express this notion in the iteration domain directly, we leverage one existing concept from the polyhedral model \cite{bastoul-habil:12}. Let $A_1,\dots,A_k$ be tables with iteration domains
$Q(A_1),\dots,Q(A_k)$ and translated self-conflict lattices $q_{A_i}+L(A_i)$. The access function mapping an iteration domain vector to an element in operand $A_i$ is  $\pi_i: Q(A_1,\dots,A_k)\to Q(A_i)$, the projection onto the $i$-th operand domain.
The set of all potential misses in $Q(A_1,\dots,A_k)$ is then 
  $$G(A_1,\dots,A_k)=\bigcup \Gamma_i(A_1,\dots,A_k)$$
  where
  \begin{equation*}
    \begin{aligned}
      \Gamma_i(A_1,\dots,A_k) & = \{ \ve x \in Q(A_1,\dots,A_k) \cap H \:: 
      \pi_i(\ve x) \in q_{A_i} + L(A_i) , i=1,\ldots,k\}\\
      &= \bigcup_{q\in q_{A_i} + L(A_i)}\mathcal{R}_i(q).      
    \end{aligned}
  \end{equation*}

\begin{figure}
  \centering
  \begin{tikzpicture}[scale=.5]
  \draw [gray,very thin] (-.5,-.5) grid (9.5,9.5);
  \fill[color=gray,opacity=.08] (0,0) rectangle (9,9);
  \draw (8.5,8.5) .. controls (8.8,9.4) and (9.2,9.9) .. (10,10.3)
  node [draw=none,anchor=west] at (10,10.3) {$Q(A,B)$};
  
  \begin{scope}
    \foreach \x in {0,...,2} {
      \foreach \y in {0,...,9} {
        \node (thisnode) at (4*\x,\y) 
        [draw,circle,minimum height=.09cm,inner sep=.02cm,fill] {};
      }
    }
    \node[draw=none,below,yshift=-.2cm] (pA0)  at (0,0) {$0$};
    \node[draw=none,below,yshift=-.2cm] (pAN)  at (4,0) {$N$};
    \node[draw=none,below,yshift=-.2cm] (pA2N) at (8,0) {$2N$};
    \draw[thick] (0,0) -- (9,0) node[draw=none,anchor= west] at (9.5,0) {$A$};
    
  \end{scope}
  \begin{scope}
    \foreach \x in {0,...,9} {
      \foreach \y in {0,...,1} {
        \node (thisnode) at (\x,4*\y+3)
        [draw,circle,minimum height=.18cm,inner sep=0pt] 
        {};
      }
    }
    \node[draw=none,left,xshift=-.2cm] (pB0)  at (0,3) {$3$};
    \node[draw=none,left,xshift=-.2cm] (pAN)  at (0,7) {$N+3$};
    \draw[thick] (0,0) -- (0,9) node[draw=none,anchor= south] at (0,9.5) {$B$};
    
  \end{scope} 
  \begin{scope}
    \foreach \x in {0,...,2} {
      \foreach \y in {0,...,1} {
        \node (thisnode) at (4*\x,4*\y+3)
        [draw,diamond,minimum height=.4cm,minimum width=.4cm,inner sep=0pt] 
        {};
      }
    }
  \end{scope} 
\end{tikzpicture}
  \caption{Illustration of conflicts for joint iteration domain of two
    vectors $A$ and $B$ with $\phi_A(0)=0\imod N$,
    $\phi_B(0)=3\imod N$, $N=4$.  Self-conflicts $G_A$ from $A$ are
    drawn as filled circles, self-conflicts $G_B$ from $B$ as hollow
    circles, cross-conflicts between points in $Q$ and
    $q_{A,B}=(0,3)^\top$ as diamonds. These cross-conflicts $\ve x$ are the
    points where $|T(\ve x)|>1$.}
  \label{fig:conflicts-and-conflict-weave}
\end{figure}
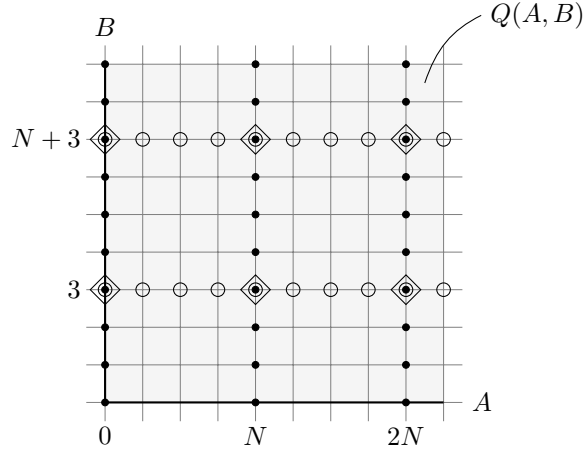

Iteration domain potential conflicts appear exactly at points in
$Q(A_1,\dots,A_k)$ where multiple operand potential conflicts meet, as can be seen geometrically in
Figure~\ref{fig:conflicts-and-conflict-weave}. Some points are included in $G$ from
multiple $G_i$ and can be declared as follows:

\begin{definition}[potential conflict index-set, potential conflict level]
  A point $x\in G$ has \emph{potential conflict index-set} 
  $$T(\ve x) =\{i\::\: x\in G_i, i\in\{1,\dots,k\}\}.$$ and \emph{potential conflict level} $\left\vert{T(\ve x)} \right\vert $
  For points $y\in Q(A_1,\dots,A_k)\setminus G(A_1,\dots,A_k)$ we set
  $T(\ve y)=\emptyset $.
\end{definition}

The set $G$ and the potential conflict set $T(\ve x)$ together completely
describe the potential conflicts in $Q(A_1,\dots,A_k)$. This machinery will be used
in subsequent sections to define, in combination with ordering information, actual cache
miss counts. 

\subsection{Actual Cache Misses}\label{sec:actual-misses}

We now want to study the geometry of sets of truly conflicting
points. As every measure of actual cache misses depends on the ordering taken, we can first describe the conditions under which no such misses arise, or alternatively for the presence of cache misses, independently of the ordering. 

\begin{definition}[Actual cache miss presence] \label{def:actual-presence}
Let $C=(c,l,K,\rho)$ be a cache specification with $N=\tfrac{c}{lK}$
  cache sets. Given a $(m_1,\dots,m_d)$-table $A$ with array $a(A)$
  and an index map $\phi$ we say that a set $S$ of points in
  $Q=[0,m_1-1]\times\dots\times[0,m_d-1]\cap\Z^d$ \emph{will contain cache misses}, if 
  $\left\{p\in S\::\: |\{x\in S\::\: \phi(x)=\phi(p)~\imod N\}|>K\right\}\neq\emptyset$,
 i.e. if it contains at least one class of more than $K$
  potentially conflicting points.
\end{definition}

Note that this definition excludes \emph{cold misses} (see section \ref{sec:cache}) since their presence is inevitable. Since $L(C,\phi$) is a lattice we can first only consider sets with potential conflicts that contain $(0,\dots,0)\in\Z^d$, as all other nonempty
sets behave similarly under a suitable translation. We later define the set of translations that complete the analysis for all sets. 

Let us consider a subset $S$ of the integral points in the box
$$[0,m_1-1]\times\dots\times[0,m_d-1]\cap\Z^d$$ versus those in 
$$[0,m_1-1]\times\dots\times[0,m_d-1]\cap L(C,\phi).$$

We know from Definition \ref{def:actual-presence} that additional (non-cold) cache conflicts will arise whenever more than $K$ lattice points are contained in $S$. The actual number of misses seen depends entirely on the ordering taken. The key measure in deciphering total miss volume is the per-set cache pressure between successive reuses of an operand's data. This concept is similar to the \emph{reuse distance} notion from basic compiler optimization \cite{ding-zhong:03}. Classical reuse distance would reveal the number of total elements loaded between successive reuses of an operand's data element. In our framework, we are interested in a subset of that number -- the number of data elements loaded into a specific cache set between successive reuses, and we are interested in the union of such subsets to give the total measure. The remainder of this section shows how such a measure is formulated. We must first embed knowledge about the ordering taken through the iteration domain. 

Let $C=(c,l,K,\rho)$ be a cache specification and $D$ the iteration
domain. Let $L(C,A_i) $ be the lattice in $\Z^{dim(A_i)}$ generated by
$C$ and operand $A_i$ (for $i\in\{1,\dots,m\}$). We extend the lattice\fxnote{maybe a small lattice primer --da}
$L(C,A_i)$ for $A_i$ by standard lattices in the other operand's index
spaces to form a $d$-dimensional lattice in $D$ by
$$\Lambda(C,A_i) = \Z^{d_1}\times \dots \times \Z^{d_{i-1}}\times
L(C,A_i)\times \Z^{d_{i+1}}\times\dots\times\Z^{d_k}.$$ Since the cache $C$ remains constant we
will refer to this simply as $\Lambda(A_i)$. We will denote the projection of a point in $\ve x\in\Lambda(C,A_i)$ onto the $i$-th component of the product by $\pi_i(\ve x)$.

The joint set of conflicts arising from the operand lattices in $D$ is given by
$$\Lambda^{D}=D\cap\bigcup_{p=1}^m\Lambda(A_p).$$

Each point in this set (which typically is not itself a lattice) represents a potential cache conflict through
at least one operand. Some points represent conflicts arising from multiple
operands simultaneously. Counting the maximum possible multiplicity at every point yields an upper
bound. But when traversing the points of $D$ in the order given by
$\prec$ we might benefit from re-use. Assuming perfect reuse, we count one instance of each operand lattice only. This represents the lower bound. Since perfect reuse is atypical, both bounds may deviate
significantly from the true number of cache conflicts. 

For more information we consider the reuse domain $\mathcal{R}_i(q)$
for a given index $q \in Q(A_i)$. Let $x \in \mathcal{R}_i(q)$ have
subsequent reuse $x' \in \mathcal{R}_i(q)$. The associated distance
function $\Delta_{L^{D}}(\ve x,\ve x')$ then describes the cache
pressure between $\ve x$ and $\ve x'$. We will count a cache miss at
$\ve x'$ unless $\Delta_{\Lambda^D}(\ve x, \ve x')$ is low enough to ensure
that during traversal of $\Lambda^D$ in the sequence prescribed by
$\prec_{\Lambda^D}$ the data indexed between $\ve x$ and $\ve x'$ cannot
have evicted $\ve x$, i.e. when $\Delta_{\Lambda^D}(\ve x,\ve x') \leq K$
and $\ve x' \in L^{D}$. Only a single miss occurs when
$\Delta_{\Lambda^D}(\ve{x},\ve{x'}) > K$, despite the extent to which the
cache set handling $\ve x$ and $\ve x'$ is overfilled.

Denote the sequence of points of $\Lambda(A_i)$ traversed in the order $\prec_{\Lambda^D}$ by $S(A_i)=(\ve x_i^0,\dots,\ve x_i^{s_i})$. Each element in this sequence can be classified either as a `miss' or a `reuse': We call $x_i^k$ a \emph{reuse point for $A_i$ in $\Lambda^D$ under $\prec_{\Lambda^D}$} if and only if
$$\{x_i^j\::\: j<k\wedge x_i^k\in\mathcal{R}_i(\pi_i(x_i^j))\wedge \Delta_{\Lambda^D}(x_i^j,x_i^k)\leq K\}\neq \emptyset,$$
i.e. if there exists a point earlier in the sequence for which $x_i^k$ corresponds to a subsequent reuse, and the traversal distance is less than $K$, the cache set associatity. If no such earlier point exists, or if the distance of all these points is too large we call $x_i^k$ a \emph{miss point}. This partitions $S(A_i)=S_{\text{miss}}(A_i)\cup S_{\text{reuse}}(A_i)$.

Given a set $J\subseteq \Lambda^{D}$ we can thus count the number of cache misses as follows: For each point in $\ve x\in J$ we consider the potentially conflicting points $T(\ve x)$ and count only those that are miss points for the respective operand:

\begin{equation}
  \label{eqn:misses}
  \#\mathrm{Misses_J}=\sum_{\ve x \in J} \sum_{p\in T(\ve x)}  (1_{S_{\text{miss}}(A_p)}(\ve x))  
\end{equation}
where $1_X(x)$ is the indicator function for $x\in X$. 

This quantity is parametric in the cache specification, the table
sizes (where padding may be allowed), the orders induced by the
operand layout $\phi_{A_i}$, but primarily by the iteration ordering
$\prec$. In this paper we will study the minimization problem subject
to changes in $\prec$.

\section{Tiling Cache Miss Model and Code Generation} \label{sec:tiling-model}

\subsection{Tiling for Cache Performance}

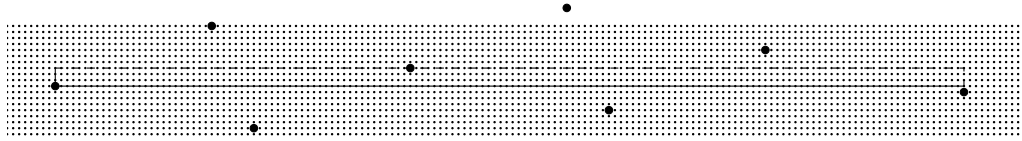
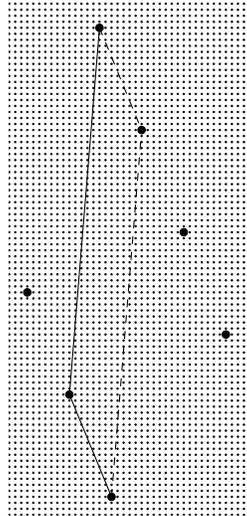
\begin{figure}
  \centering
  \subfigure[][Largest (half-open) rectangle with only 1 interior lattice point
  has volume 453 (see~{\cite[$A_7$ in Tab. IV]{ghosh-martonosi-malik:99}}). It is too large to be used in a regular tiling.]{%
    \label{fig:rectangles:a}%
    \begin{tikzpicture}[xscale=0.08,yscale=0.08]
  \begin{scope}
    \clip (-8,-10) rectangle (161,15); 
    
    \foreach \i in {-8,-7,...,160}{
      \foreach \j in {-8,-7,...,10}{
        \draw ($\i*(1,0)+\j*(0,1)$) 
        node[draw,fill,circle,minimum size=.5pt,inner sep=0pt] {};
      }
    }
    \draw (151,0) -- (0,0) -- (0,3);
    \draw[dashed] (0,3) -- (151,3) -- (151,0);
    
    \pgftransformcm{5}{61}{7}{-17}{\pgfpoint{0cm}{0cm}}
    \foreach \i in {-20,-19,...,5}{
      \foreach \j in {-25,-24,...,25}{
        \draw ($\i*(1,0)+\j*(0,1)$) node[draw,fill,circle,inner sep=1pt] {};
      }
    }
  \end{scope}
\end{tikzpicture}
%
  }
  \hfill
  \subfigure[][{Fundamental region of lattice has volume
  $\left|\det\begin{pmatrix}
    5 & 7\\
    61 & -17\\
  \end{pmatrix}\right|=512$}]{%
    \label{fig:rectangles:b}%
    \begin{tikzpicture}[xscale=0.08,yscale=0.08]
  \begin{scope}
    \clip (-10,-20) rectangle (30,70); 
    
    \foreach \i in {-10,-9,...,30}{
      \foreach \j in {-20,-19,...,65}{
        \draw ($\i*(1,0)+\j*(0,1)$) 
        node[draw,fill,circle,minimum size=.5pt,inner sep=0pt] {};
      }
    }
    \draw (0,0) -- (5,61);
    \draw (0,0) -- (7,-17);
    \draw[dashed] (7,-17) -- +(5,61);
    \draw[dashed] (5,61) -- +(7,-17);
    
    \pgftransformcm{5}{61}{7}{-17}{\pgfpoint{0cm}{0cm}}

    \foreach \i in {-16,-15,...,16}{
      \foreach \j in {-20,-19,...,20}{
        \draw ($\i*(1,0)+\j*(0,1)$) node[draw,fill,circle,inner sep=1pt] {};
      }
    }
  \end{scope}
\end{tikzpicture}
%
  }
  \caption[Tile volume difference between rectangular and lattice
    tiling.]{Tile volume difference between rectangular and lattice
    tiling. For illustration we use the lattice
    of~{\protect\cite[Fig. 14]{ghosh-martonosi-malik:99}}, generated by
    \ensuremath{\begin{pmatrix}
        5 & 7\\
        61 & -17\\
      \end{pmatrix}}. 
    Note furthermore that if a rectangular tiling
    is constructed from a scaled copy of the rectangle containing more
    than one lattice point the number of integral points can vary
    across tiles, in a lattice tiling~\subref{fig:rectangles:b} it is
    constant (except at the boundary). (To simplify the picture
    instead of all integer points only those whose coordinates are
    divisible by $10$, and the figure has been scaled in a $1:2$ ratio
    in the $y$-direction.)  }
  \label{fig:rectangles}
\end{figure}

In  Section \ref{sec:cache} we made some informal observations about cache capacity and described the need to view cache models from the perspective of one cache-set. When considering iteration space tiling, this effect is particularly important. While much important work has been performed on iteration space tiling both using traditional compiler optimizations and in the polyhedral model, (as described in Section \ref{sec:related-work}, selection of best-performing tile size and shape has remained more of an art than science, an so called auto-tuning solutions are often relied-upon to compute the best tile. Nowhere in the literature are tiles shaped and sized according to the natural structure imposed by the hardware. In this section we we describe how tiles that are constructed according to the cache's natural associativity lattice exhibit two clear theoretical advantages.

Let $A_1,\dots,A_k$ be operands with index-sets $Q(A_i)$, index-maps
$\phi_{A_i}$ and iteration domain $Q(A_1,\dots,A_k) \cap H$. Let
$C=(c,l,k,\rho)$ be a cache specification, and let the lattice
$L(C,A_i)$ describe potential cache misses. Consider vectors
$(\ve l_1,\dots,\ve l_{m_i})$ generating the operand lattice
$L(C,A_i)$. Typical lattice generators are not aligned with the matrix
dimensions, i.e. not simply integral multiples of unit vectors. Hence
rectangular regions of the operand index set $Q(A_i)$ will have an
unpredictable number of lattice points contained within as illustrated
in Figure~\ref{fig:rectangles:a}. Let us instead consider half-open
parallelepiped tilings that are formed from integral multiples of the
lattice basis vectors. An example of such tilings is shown in
Figure~\ref{fig:rectangles:b}. There are two distinct theoretical
advantages to such lattice tilings over the rectangular form. Firstly,
adjacent tiles of tile (a) contain a variable number of lattice points
and will therefore induce a non-constant number of cache misses. The
implications of this from a cache perspective may be severe. Even if
one rectangular tile was sized very carefully to minimize the number
of cache misses, adjacent tiles may have a completely different cache
miss behavior. This, combined with the community's focus on rectangular tilings may
explain why it has proven so difficult to provide clear guidance on tile
size/shape. Adjacent tiles of type (b) contain identical numbers of
lattice points, as long as the tiles are whole. Secondly, the volume
of tiles of type (b) can easily be as much as 20\% larger than tiles
containing the same number of lattice points but of type (a). Even in the small example of~\cite[Fig. 14]{ghosh-martonosi-malik:99} the
best rectangular volume is $453$, the one chosen by the authors has volume $416$, while the parallelepiped volume of type (b)
is given by $\left|\det\begin{pmatrix}
    5 & 7\\
    61 & -17\\
  \end{pmatrix}\right|=512$, a saving of 13\% resp. 24\%. Greater tile volume
can directly lead to greater performance since tile boundaries
typically enforce cache misses.

\subsection{Tiling Mechanics}\label{sec:tiling}

We follow the tiling methodology of~\cite{goumas-athanasaki-koziris:03}:
A tile is the half-open parallelepiped generated by linearly
independent vectors $\{\vec p_1,\dots,\vec p_d\}$, so that the matrix
$(\vec p_1 \cdots \vec p_d)$ bijectively identifies the unit cube with
a tile, and the lattice points of the 
standard lattice $\Z^d$ correspond to the footpoints of the
tiles\fxnote{picture}. Given some iteration domain $D\subset\Z^d$ and\fxnote{H is now used earlier}
writing $\mat H=(\vec p_1 \cdots \vec p_d)^{-1}$ the prototypical tile
(or \emph{single-tile iteration space}) starting at the origin is
thus given by
\begin{equation}
  \label{eq:tile}
  P_D(\mat H)= D\cap\{\vec x\in\Z^d | \vec 0\leq \vec x < \vec p_i, i\in\{1,\dots,d\}\}
        = \{\vec x\in D | \vec 0\leq \lfloor \mat H\vec x \rfloor <
        \vec 1\},
\end{equation}
while the set of footpoints (or \emph{tile iteration space}), i.e.,
the translation vertors to cover $D$ with tiles $P_D(H)$is
\begin{equation}
  \label{eq:footpoints}
  T_D(\mat H)=\left\{\vec t | (\vec p_1 \cdots \vec p_d)\vec t+P_D(\mat H)\cap D\neq\emptyset\right\}
        =\left\{\vec t\in \Z^d | \vec t=\lfloor \mat H \vec x\rfloor,
          \vec x\in D\right\}.
\end{equation}
This makes $D\subseteq P_D(\ma H)+\ma H^{-1}T_D(\ma H)$, i.e., the Minkowski sum of
$P_D(H)$ and $\ma H^{-1}T_D(\ma H)$ is a covering of $D$. 

When $D$ and $\mat H$ are clear from the context we will simply write $P$
for the tile and $T$ for the translations. Abusing notation we write
$\vec x\in P^{\vec t}$ to denote $\vec x\in \ma H^{-1}\vec t+P$, a point in the
affine translate of $P$ by a transformed vector $\vec t\in T$

The tiling transformation can be written as 
$$r:\Z^d\to \Z^{2d},\quad r(\vec x)=
\begin{pmatrix}
  \lfloor \mat H\vec x\rfloor\\
  \vec x-\mat H^{-1}\lfloor \mat H\vec x\rfloor
\end{pmatrix},$$
i.e., $r$ assigns to each point $\vec x$ the respective
footpoint in $T_D(\mat H)$ in the first $d$ coordinates, and the point
inside the tile $P_D(\mat H)$ in the second $d$ coordinates.

\subsection{Cache Tiling Model}

The cache miss counting scheme described in Section~\ref{sec:actual-misses} extends easily to the tiled case.  We want to first consider the joint set of conflicts arising from the operand lattices in the tile $P^{\ve t}=\ma H^{-1}\vec t+P$ rather than in $D$, which is given by
$$\Lambda^{\ve t}=P^{\ve t}\cap\bigcup_{p=1}^m\Lambda(A_p).$$

We will then consider the reuse domain $\mathcal{R}_i(q)$ for a given index\fxnote{check which $i$ --uuh} $q \in P^{\ve t}$. The analysis regarding subsequent points in $P^{t}$ is indentical to that for $D$, and so we 
define the sequence of points of $\Lambda(A_i)$ traversed in the order $\prec_{\Lambda^{\ve t}}$ by $S^{\ve t}(A_i)=(\ve x_i^0,\dots,\ve x_i^{s_i})$, and partition it into miss and reuse as $S^{\ve t}(A_i)=S^{\ve t}_{\text{miss}}(A_i)\cup S^{\ve t}_{\text{reuse}}(A_i)$ like before and restate Equation~\eqref{eqn:misses} for the tiled case as

\begin{equation}
  \label{eqn:tiled-misses}
  \#\mathrm{Misses_J}=\sum_{\ve x \in J} \sum_{p\in T(\ve x)}  (1_{S^{\ve t}_{\text{miss}}(A_p)}(\ve x))
\end{equation}
for every $J\subseteq \Lambda^{\ve t}$.

\section{Computational Results}\label{sec:experimental}
In this section we describe computation results of using the
associativity lattice tiling framework to accelerate matrix
multiplication programs. 
We do not describe the implementation in
any detail, except to say that it is a C++ framework that uses the
packages ClooG~\cite{cloog} for loop bound generation and
NTL~\cite{NTL} for integer math library support. The implementation
currently can only perform tiling of matrix multiplication, along with
associated solution checking against a library implementation
(BLAS). In future work we will extend this framework to further dense
linear algebra and to non matrix operations.
 
From a set of problem specifications, array layout characteristics,
pointers to memory locations, padded dimensions etc, the
implementation generates the associativity latices of the operations
and using equation~\eqref{eqn:misses}
or~\eqref{eqn:tiled-misses} constructs the
appropriate cache miss model. The best in a small search of tiling
options is chosen and the the code generated using the tiling
specification from section~\ref{sec:tiling} and using the CLooG
library.


\subsubsection{Lattice Tilings versus Compiler Optimization}

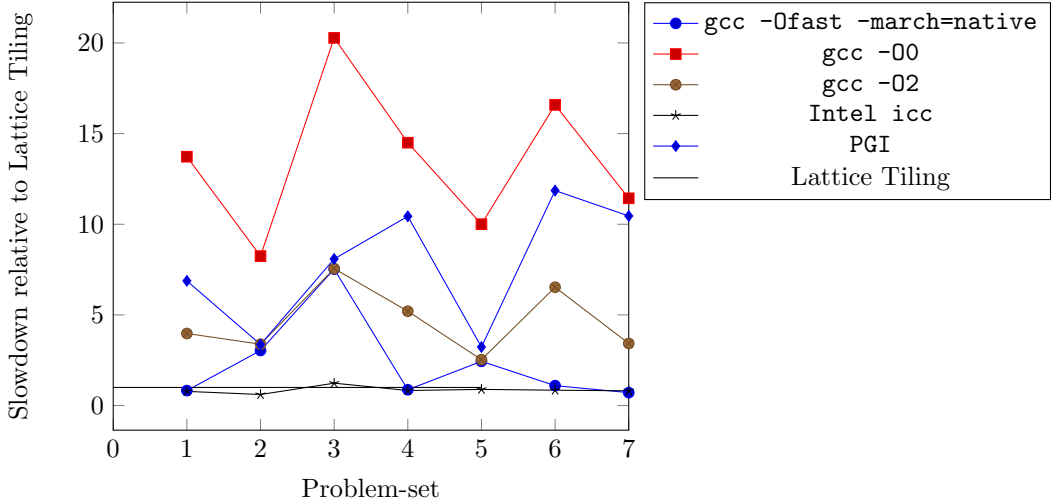
\begin{figure}
  \centering
 \begin{tikzpicture}
\begin{axis}[legend entries={ \texttt{gcc -Ofast -march=native},\texttt{gcc -O0},\texttt{gcc -O2},\texttt{Intel icc},\texttt{PGI}, Lattice Tiling},
legend, legend pos=outer north east,xlabel=Problem-set,
    ylabel=Slowdown relative to Lattice Tiling, xmin=0,xmax=7]
\addplot table [x index=0, y index=1]{\mytable};
\addplot table [x index=0, y index=2]{\mytable};
\addplot table [x index=0, y index=3]{\mytable};
\addplot table [x index=0, y index=4]{\mytable};
\addplot table [x index=0, y index=5]{\mytable};
\addplot[mark=none, black] {1.0};
\end{axis}
\end{tikzpicture}
\caption{Computational Results of Tiling Using Associativity Lattices against aggressive compiler optimizations for \texttt{gcc} (5.1.0), \texttt{gcc-graphite} (5.1.0), Intel \texttt{icc} (15.0.3), and \texttt{pgi} (15.10-0).}
\label{fig:lattice-tiles-compilers}
\end{figure}

Figure~\ref{fig:lattice-tiles-compilers} shows computational results
of tiling using the associativity lattices against various compilers
and flags. In general, the lattice tiling performs much better than
expected. Some compilers, such as \texttt{pgi} do not seem to be able to enable
cache tiling for this problem and their performance is many times
slower than our lattice tiling. Compared to unoptimized code
(\texttt{gcc -O0}), the lattice tiling produces a speed-up of 10 to
20. Against the more typical optimization level \texttt{gcc -O2}, our
framework gives a speed up of around 2-6 times. Our framework
sometimes gives no advantage over \texttt{gcc} with aggressive
optimization while for other problems, we see a clear 2-3 times
speedup. The Intel compiler with aggressive optimization is able to
tile for cache usually as well as the lattice tiling.

We consider the performance improvements from lattice tiling
surprising because currently we only tile for a single level of the
memory hierarchy. These results were obtained by tiling for L1 cache
on the Intel Haswell Architecture. In future work, we will present
results for multiple levels of tiling.

\subsubsection{Rectangular versus Lattice Tilings}

\begin{figure}
  \centering
  \begin{tikzpicture}[scale=.25,line join=bevel,z=-10]

\draw [dotted] grid (15,15);

\coordinate (origin) at (0,0);

\draw (origin) -- (15,0) -- (15,15) -- (0,15) -- cycle;

\foreach \t in {(0,0)}{
  \begin{scope}{shift=(\t)}
    \draw (0,0) -- (9,1);
    \draw [dashed] (9,1) -- (10,9) -- (1,8);
    \draw (1,8) -- (0,0);
  \end{scope}
}

\foreach \a in {0,1,...,36}{
  \draw [very thick] 
  let \p1=({int(mod(7*\a,16))},{int(div(7*\a,16))})
  in (\p1) node[star, inner sep=1pt, draw=green, fill=green] {};
}
\foreach \a in {(0,0), (5,1), (3,2), (8,3), (6,4), (4,5), (9,6), (7,7),
(5,8),
(1,3), (2,6)}{
  \draw [green,very thick] \a -- +(6,0);
}
\end{tikzpicture}
  \caption{Reduced spatial reuse of lattice tiles compared to
    rectangular tiles: cache lines starting inside the fundamental
    region may not be used completely for the respective tile, and the
    adjacent tiles may not reuse all entries loaded for some operand
    due to non-orthogonal shifts from tile to tile.}
  \label{fig:spatial}
\end{figure}
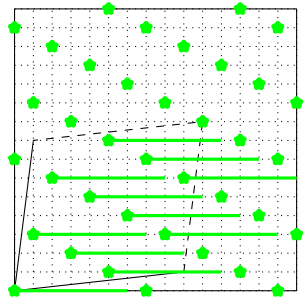

Figure~\ref{fig:lattice-tiles-compilers} compares the results of best
rectangular tilings versus best lattice tilings. Although we have
described clear theoretical advantages of lattice tiles versus
rectangular tiling, the two methods appear quite close in
performance. The reason for this is that while lattice tiles improve
addressable volume they also display worse spatial reuse
characteristics, as shown in Figure~\ref{fig:spatial}. 
Given the preliminary state of our implementation we believe that a
significant advantage of the lattice tiling for certain problem sizes
can be exhibited with a properly tuned code generation procedure as
implemented by the compiler optimization passes for rectangular tiles.

\subsubsection{Auto-Threading}

\begin{figure}
  \centering
 \begin{tikzpicture}
\begin{axis}[legend entries={Lattice Tiling, \texttt{gcc-graphite}},
legend, legend pos=outer north east,xlabel=Threads,
    ylabel=Execution Time]
\addplot table [x=Size,y=1]{\mytables};
\addplot table [x=Size,y=2]{\mytables};
\end{axis}
\end{tikzpicture}
 \caption{Computational Results of Automatic openMP Threading Using Associativity Lattices and with \texttt{gcc-graphite}.}
\label{fig:autothread}
\end{figure}
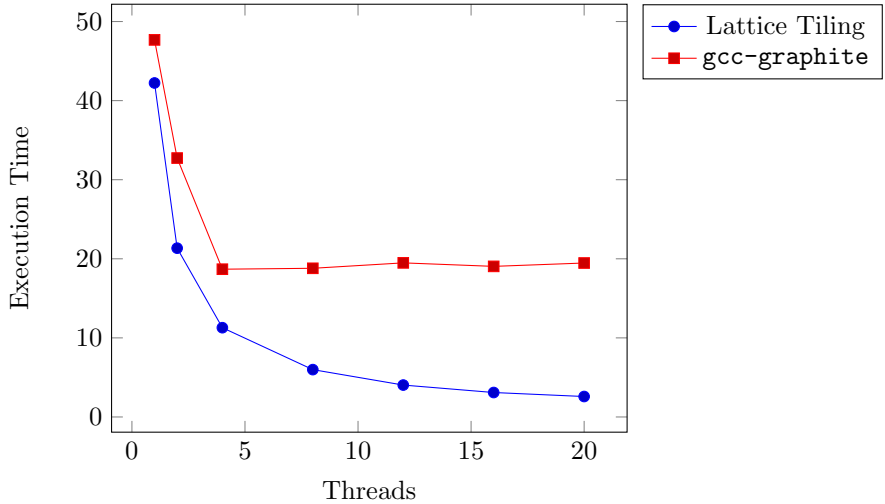

Our tiling implementation also displays basic automatic
parallelization capabilities using openMP directives. The speed-up of
our generated codes versus \texttt{gcc-graphite}, another
automatically parallelizing polyhedral compiler, are shown in
Figure~\ref{fig:autothread}. For the problem size chosen, our
framework was able to generate parallel codes that exhibit speed-up on
20 threads of 20 Intel Haswell cores. On the other hand, \texttt{gcc-graphite} was able to produce
speed-ups only up to 4 threads.

\subsubsection{Analysis/Model Cost}

The cost of direct evaluation of Equation~\eqref{eqn:tiled-misses} is exponential and thus an efficient implementation is not possible. This should not be a surprise, since Equation~\eqref{eqn:tiled-misses} precisely evaluates every case miss in the code region of the iteration domain, and full evaluation of that requires a similar number of data accesses to the code being modeled. In its raw form then, Equation~\eqref{eqn:tiled-misses} is no improvement over the CME approach described in~\cite{ghosh1997cache} and could not be incorporated into a compile-time or time optimization framework. Our approach however, unlike the CME approach leads to a solution set with a clear structure that can be exploited in many ways, making it unnecessary to generate the full, intractable solution set. Equation~\eqref{eqn:tiled-misses} is evaluated for every tile in the domain and for every set in the cache. An improvement to our method would model a few certain tiles for a few certain sets using a sampling approach. The details regarding which tiles and sets to select are sufficient to fill a follow-on paper. However, we believe that sufficient structure is expressed using the lattice framework, and that sufficiently robust mathematical toolsets exist such that only a small fraction of the analysis in~\eqref{eqn:tiled-misses} can be actually evaluated, while still providing reliable approximate cache miss information about the whole code. 

We also here describe an alternative approach, that uses a common-sense tiling mechanism rather than being driven by cache miss modeling. Since lattice tiles can be constructed without counting lattice points explicitly, then a sensible tiling can be constructed by choosing a known number of lattice points to be contained from the operand that is tiled by lattices. We would choose the number of lattice points in the tile to be in the range $[K-\alpha,K+\beta]$. We would expect that $\beta=0$ since traversing more than $K$ points in the operand's tile would mean that no reuse could occur between tile slices. We would expect that $\alpha$ would be  small, e.g. $\alpha=2$ or $\alpha=1$. Experimentally we have observed that lattice tiles that contain $K-1$ lattice points perform well and so for our experiment this is chosen. The remaining operands will be tiled rectangularly with sizes induced by the first operand's lattice tile shape. The cost of this tiling analysis is dominated by lattice basis reduction using the NTL library, which is not significant for the low dimensional lattices and the (possibly padded) matrix dimensions appearing, which are often powers of two
. \fxnote{need results here} We envisage a hybrid approach where direct analysis is performed and a small search of modeled tiles is evaluated to decide which of the small sample set is optimal. 



\bibliographystyle{amsalpha}
\bibliography{cache}

\end{document}